\pdfoutput=1
\documentclass[11pt]{article}
\usepackage{jheppub}
\usepackage[all]{xy}
\usepackage{customprelude}

\usepackage{tabu}
\usepackage{multirow}

\newcommand\be{\begin{equation}}
\newcommand\ee{\end{equation}}

\newcommand\CY{\mathbf{X}_6}
\newcommand\TX{{\mathcal T}_{\CY}}
\newcommand\Tgg{\cT[\fg,\fg']}

\title{\centering Higher Form Symmetries of Argyres-Douglas Theories}

\author[\sharp \ast]{Michele Del Zotto,}
\author[\sharp]{Iñaki García Etxebarria,}
\author[\sharp]{Saghar S. Hosseini}

\affiliation[\sharp]{Department of Mathematical Sciences\\
Durham University, Durham, DH1 3LE, United Kingdom}
\affiliation[\ast]{Department of Mathematics, and Department of Physics and Astronomy\\ Uppsala University, Uppsala, Sweden}

\emailAdd{michele.delzotto@math.uu.se}
\emailAdd{inaki.garcia-etxebarria@durham.ac.uk}
\emailAdd{sagharsadat.hosseinisemnani@durham.ac.uk} 

\abstract{We determine the structure of 1-form symmetries for all 4d
  $\mathcal{N}=2$ theories that have a geometric engineering in terms
  of type IIB string theory on isolated hypersurface
  singularities. This is a large class of models, that includes
  Argyres-Douglas theories and many others. Despite the lack of known
  gauge theory descriptions for most such theories, we find that the
  spectrum of 1-form symmetries can be obtained via a careful analysis
  of the non-commutative behaviour of RR fluxes at infinity in the IIB
  setup. The final result admits a very compact field theoretical
  reformulation in terms of the BPS quiver. We illustrate our methods
  in detail in the case of the $(\mathfrak{g},\mathfrak{g}')$
  Argyres-Douglas theories found by Cecotti-Neitzke-Vafa. In those cases where $\cN=1$ gauge theory
  descriptions have been proposed for theories within this class, we
  find agreement between the 1-form symmetries of such $\cN=1$
  Lagrangian flows and those of the actual Argyres-Douglas fixed
  points, thus giving a consistency check for these proposals.}

\begin{document}


\maketitle

\section{Introduction}

Four-dimensional supersymmetric quantum field theories are an ideal
laboratory to probe our understanding of strongly coupled relativistic
systems. In this context, $\cN=2$ supersymmetric models are
particularly useful since they have enough supercharges to constrain
the perturbative corrections while still allowing for interesting
dynamics. In the past two decades a wide variety of new $\cN=2$
theories have been constructed exploiting various kinds of geometric
engineering techniques in string theory --- for a nice review see
\cite{Tachikawa:2013kta}.
There is no known Lagrangian description for many of the $\cN=2$
theories so obtained, which makes it essential to develop techniques
to study these theories that are independent from a Lagrangian
formulation.
 A natural basic question in this context is to determine all
the symmetries for these models, including the higher symmetries
\cite{Gaiotto:2014kfa}. In this paper we take a first step in this
direction, and determine the 1-form symmetries of a broad class of
$\cN=2$ theories in four dimensions, namely those that arise from IIB
on hypersurface singularities. Many well known $\cN=2$ theories belong
to this class, and we will analyse the details of a number of examples
below.

\medskip

One generic feature of the theories that we study is that non-local
BPS dyons become simultaneously massless.  The original examples of
this phenomenon were given by Argyres and Douglas in
\cite{Argyres:1995jj,Argyres:1995xn}. Since the massless degrees of freedom are mutually
non-local, the corresponding dynamics cannot be described by a
conventional Lagrangian.\footnote{   At least at the IR fixed point, but
  this does not exclude the existence of non-conformal Lagrangian
  theories in the same universality class.
  We will study some examples
  of this phenomenon below.} Moreover, by the scale invariance of the
corresponding Seiberg-Witten (SW) geometry, the theories are argued to be
$\cN=2$ superconformal.

\medskip

Argyres-Douglas theories can be realized in Type IIB superstrings on
isolated hypersurface singularities
\cite{Eguchi:1996vu,Eguchi:1996ds}. This perspective allows to compute
the corresponding spectrum of BPS states from the bound states of D3
branes on vanishing special Lagrangian 3-cycles
\cite{Shapere:1999xr}. The same geometric construction can be
generalized to more general hypersurface singularities at finite
distance in moduli space\footnote{   See \cite{Cecotti:2010fi,Cecotti:2011gu,DelZotto:2011an,Cecotti:2012jx,Xie:2012hs,Cecotti:2013lda,DelZotto:2015rca,Xie:2015rpa,Giacomelli:2017ckh} for more examples of this kind, as well as \cite{Chen:2016bzh,Chen:2017wkw} for interesting generalizations beyond the class of hypersurface singularities.}  --- which translates to the requirement that
the corresponding 2d (2,2) Landau-Ginzburg worldsheet theory has
central charge $\hat c < 2$ \cite{Gukov:1999ya,Shapere:1999xr}. 
Each such model is characterized by a quasihomogeneous polynomial
\be\label{eq:scaling}
f( \lambda^{w_i} X_i) = \lambda^d f(X_i)
\ee
where $d$ and $w_i$ are positive integers known respectively as degree and weights of the singularity, and the corresponding geometry is given by $\CY \df \{f(X_i)=0\}$.  The $\mathbb C^*$ action in \eqref{eq:scaling} plays the same role of the scale invariance for the SW geometry: it is well-known that this can be exploited to compute the dimensions of the various Coulomb branch operators of the SCFT \cite{Shapere:1999xr}. These singularities are at finite distance in moduli space 
provided $\sum_i  w_i > d$ which is the singularity theory translation of the condition $\hat c <2$ \cite{Vafa:1988uu,Lerche:1989uy}. 

\medskip

The interesting part of the defect group for theories of this class is given by\footnote{  The definition of defect group we use in this paper is \cite{Albertini:2020mdx}
\be\label{eq:defecto}
\mathbb D \df \bigoplus_{n} \mathbb D^{(n)} \qquad\text{where}\quad  \mathbb D^{(n)}\df \bigoplus_{p \text{ branes and } k\text{ cycles} \atop \text{such that } p-k+1 = n} \left({H_k(\CY,\partial \CY) \over H_k(\CY)}\right)\,.
\ee
Notice that in the current paper we are including the non-torsional part in the definition of $\mathbb D^{(j)}$.}
\be\label{eq:generaldefecto}
\mathbb D^{(1)} = \mathbb Z^{\kappa} \, \oplus {\bigoplus}_{i=1}^n \Big( \mathbb Z_{m_i} \oplus  \mathbb Z_{m_i}\Big)
\ee
where $n$ is the rank of the SCFT and $\kappa$ is the rank of its flavor symmetry group. The $m_i$'s are positive integers that we determine below --- if for some $i$ the corresponding $m_i $ equals 1, the corresponding factor is trivial and the summand is dropped above. The torsional groups in parenthesis are non-trivially paired, meaning that the corresponding charge operators form a non-commuting Heisenberg algebra. 

\medskip

Notice that in the formulas above, we include the free factors of the
defect group. We stress here that to each free $\mathbb Z$ factor in
the $j$-form defect group $\mathbb D^{(j)}$ there is a corresponding
abelian $U(1)$ higher form symmetry which is shifted in degree by one.\footnote{  This is ultimately related to the fact that the
  pairing for the free part of the defect group is the Poincar\'e
  pairing and not the linking pairing, and these are shifted by one in
  degree \cite{ADZGEH}.} For instance, the free factor
of $\mathbb D^{(1)}$ in equation \eqref{eq:generaldefecto} corresponds to the continuous zero-form symmetries of these SCFTs. For such zero-form symmetries this group can enhance to become
non-Abelian --- see \cite{Caorsi:2018ahl,Caorsi:2019vex} for
conditions about the enhancement of the flavor symmetries in terms of
the corresponding categories of BPS states. It is also interesting to remark that in the language of those papers, the
Grothendieck group of the cluster category associated to the BPS quiver precisely coincides with
the full $\mathbb D^{(1)}$ we compute in this paper, referred to as the 't
Hooft group in \cite{CDZ,Caorsi:2016ebt} --- see \S \ref{sec:sgroto} below.

\medskip

Exploiting the Milnor-Orlik conjecture \cite{OM}, and its subsequent proof by Boyer, Galicki and Simanca for the case of interest in this paper \cite{Boyer}, we can rewrite \eqref{eq:generaldefecto} as follows:
\be
\mathbb D^{(1)} = \mathbb Z^\kappa \oplus \bigoplus_{i=1}^4\mathbb{Z}_{r_i}^{2g_i}
\ee
In terms of singularity theory data, we have that $\kappa$ is given by \cite{sasakianBIBLE}
\be
\kappa = \sum (-1)^{4-s} {u_{i_1} \cdots u_{i_s} \over v_{i_1} \cdots v_{i_s} \, \text{lcm}(u_{i_1}\cdots u_{i_s})}  
\ee
where
\be
u_i = {d \over \text{gcd}(d,w_i)} \qquad v_i = {w_i \over \text{gcd}(d,w_i)} 
\ee
and the sum is taken over all the 16 subsets $\{i_1,...,i_s\}$ with $s$ elements of the index set $\{1,2,3,4\}$. Moreover,
\be
r_i=\text{gcd}(w_1,..,\hat{w}_i,...,w_4)
\ee
and
\be
2g_i=-1+\sum_{j\neq i}\frac{\text{gcd}(d,w_j)}{w_j}-
d\sum_{j< k \atop j,k\neq i}\frac{\text{gcd}(w_j,w_k)}{w_jw_k}+d^2 \frac{r_i}{w_1...\hat{w}_i...w_4}\, ,
\end{equation}
where the notation $\hat{w}$ means that $w$ is omitted.

\medskip

Within the class of theories engineered by type IIB superstrings on hypersurface singularities, a subset of geometries that naturally generalizes the
original examples by Argyres and Douglas are the Cecotti-Neitzke-Vafa $(\fg,\fg')$ SCFTs, or $\Tgg$ for short \cite{Cecotti:2010fi}. 
These are also known as the generalized Argyres-Douglas theories of type $(\fg,\fg')$ \cite{Xie:2012hs}. Consider a background
of the form $\cM_4\times \CY$, where $\cM_4$ is some arbitrary closed
four-manifold, which we will always assume to be closed \Spin\ without
torsion, and $\CY$ is a non-compact Calabi-Yau threefold with an
isolated singularity, given by the hypersurface
\begin{equation}
  \label{eq:Argyres-Douglas-geometry}
  P_{(\fg,\fg')}(x,y,w,z) = P_\fg(x,y) + P_{\fg'}(w,z) = 0
\end{equation}
inside $\bC^4$. Here 
\begin{equation}
  \begin{array}{c|c}
    \fg & P_\fg(x,y) \\
    \hline
    A_n & x^2 + y^{n+1}\\
    D_n & x^2y + y^{n-1}\\
    E_6 & x^3 + y^4\\
    E_7 & x^3 + xy^3\\
    E_8 & x^3 + y^5
  \end{array}
\end{equation}
are such that $z^2 + P_\fg(x,y) = 0$ is the du Val singularity of type $\mathfrak g$. This space has an isolated singularity at
$x=y=z=w=0$. The corresponding 2d $(2,2)$ LG theory has superpotential $W = P_{(\fg,\fg')}(x,y,w,z)$. 
 The resulting 2d worldsheet theory has central charge $\hat c < 2$ in this case \cite{Cecotti:2010fi}, and these singularities are at finite distance. It is believed that at low energies this configuration
can be described by the $\Tgg$ four dimensional SCFT compactified on
$\cM_4$.
The non-trivial local
degrees of freedom of such SCFT arise from massless D3 branes
wrapped on the vanishing three-cycles at the singular point of $\CY$.

\medskip

The local degrees of freedom of $\Tgg$ are in this way fully
determined by the choice of geometry, but in presence of a nontrivial
defect group at the horizon of the IIB compactification, extra
information is required to fully specify the theory and its partition
function on $\cM_4$ \cite{Garcia-Etxebarria:2019cnb} --- see also
\cite{DelZotto:2015isa,Albertini:2020mdx,Morrison:2020ool}.  The most
well known example of this fact is the case of the $\cN=4$ theory with
simple ADE algebra $\fg$ corresponding to IIB on
$\mathbb T^2 \times \mathbb C^2 / \Gamma_{\fg}$. While giving
$\Gamma_{\fg}$ fully specifies the spectrum of local operators (and
their correlators), in order to compute the partition function on
topologically non-trivial manifolds we need to specify additional
data, fixing the global structure of the theory and, in particular,
its 1-form symmetries.\footnote{   In the case of $\cN=4$, the choice of
  1-form symmetries can be alternatively described as a choice of
  global form for the gauge group together with some additional
  discrete data \cite{Aharony:2013hda}, but we will avoid that
  terminology since in most cases in this paper no gauge theory
  description is known.} Our main result in this paper is to determine
the defect group for the geometries associated to the $\Tgg$ theories,
and the corresponding Heisenberg algebra of noncommuting fluxes. We
find that many of these theories admit different inequivalent global
structures and hence distinct partition functions on four-manifolds
with nontrivial intersecting 2-cycles.\footnote{  See
  \cite{Gukov:2017zao,Moore:2017cmm,Kozcaz:2018usv} for work on the
  interplay between partition functions of various $\cN=2$ SCFTs and
  the theory of 4-manifold invariants.} While this does not affect the
superconformal index for these models
\cite{Buican:2015ina,Cordova:2015nma,Cordova:2016uwk,Buican:2015tda,Cecotti:2015lab,Xie:2016evu,Xie:2019zlb}
which corresponds to the partition function on $S^1 \times S^3$, it
does affect the lens space index or more complicated partition
functions (see
e.g. \cite{Razamat:2013jxa,Razamat:2014pta,Festuccia:2016gul,Festuccia:2018rew} for some interesting examples of $\cN=2$ backgrounds that would be interesting to couple to AD SCFTs).
\medskip

As there are no known Lagrangian descriptions of most $\Tgg$ SCFTs
(with the exception of some cases; we will come back to these
momentarily) it might seem hard to find the 1-form symmetries for
these theories using purely field theoretical tools.  However, as we
explain below --- extending previous results
\cite{DelZotto:2015isa,Garcia-Etxebarria:2019cnb,Morrison:2020ool,Albertini:2020mdx}
to the four dimensional setting --- there is a way of rephrasing the
results from the IIB analysis in purely field theoretical terms. We
find that the role played by the unscreened part of the center of the
gauge group\footnote{  We will review the relevant parts of the
  construction of \cite{Aharony:2013hda} in \S\ref{sec:Maruyoshi-Song}
  below.} in the analysis in \cite{Aharony:2013hda} is played in the
non-Lagrangian setting in this paper by $\Tor(\coker \sQ)$, with $\sQ$
the BPS quiver for the theory \cite{Cecotti:2011rv,Alim:2011kw}. The free part of the group $\coker \sQ$ coincides with the factor $\mathbb Z^\kappa$, hence $\mathbb D^{(1)} = \coker \sQ$:
 it is natural to expect that this field theoretical formulation will be
general, even in the absence of a simple IIB construction, and indeed this follows from the analysis done in \cite{Caorsi:2016ebt}; we will
explore this point further in upcoming work
\cite{ADZGEH}.

\medskip

The results in this paper can be used to provide a subtle but powerful
test of any proposed Lagrangian dual description of the $\Tgg$
theories, imposing matching of 1-form symmetries. Such Lagrangian
descriptions have in fact been recently proposed by
\cite{Maruyoshi:2016aim,Maruyoshi:2016tqk,Agarwal:2017roi} (see also
\cite{Benvenuti:2017bpg,Giacomelli:2017ckh, Carta:2020plx}), in terms
of $\cN=2$ theories perturbed by $\cN=1$ relevant deformations. We
find that the Lagrangians proposed in the literature have precisely
the same structure of 1-form symmetries as the $\Tgg$ theories to
which they are believed to flow. Our computations also provide
interesting additional constraints on the existence of potential
Lagrangian descriptions for those theories for which no Lagrangians
are known.

\medskip

The structure of this paper is as follows. In section
\ref{sec:jumanji} we review the results of
\cite{Garcia-Etxebarria:2019cnb,Albertini:2020mdx} that are needed for
our analysis, as well as some aspects of the geometric engineering of
IIB string theory on hypersurface singularities. In section
\ref{sec:GGPRIME} the main result of the paper is derived (our readers
interested mainly in the result of the analysis can skip to table
\ref{table:Argyres-Douglas}), and we explain how our results are
consistent with the $\cN=1$ Lagrangian flows proposed in
\cite{Maruyoshi:2016aim,Maruyoshi:2016tqk,Agarwal:2017roi}. A 
computation in K-theory necessary to justify our mathematical
description of the defect groups is presented in the appendix.

\medskip

\emph{Note added: while this work was reaching completion, we received \cite{SAKURA2} that has a small overlap with some of our results. 
We thank the authors of that paper to accept coordinating the submission of our papers to the arXiv.}

\section{Geometric engineering and higher symmetries}\label{sec:jumanji} 

In this section we review how to compute the spectrum of higher form
symmetries for geometrically engineered $\cN=2$ theories. We refer the
reader to \cite{Garcia-Etxebarria:2019cnb} for a more in-depth
discussion, here we will just highlight the modifications necessary
when dealing with Calabi-Yau threefolds.

\subsection{Global structure and flux non-commutativity}

\label{sec:flux-noncommutativity}

Consider IIB superstring theory on 
\be
\cM_4\times \CY \,,
\ee
where $\CY$ is a local Calabi-Yau threefold. Via geometric engineering, this computes the partition function of a 4d $\cN=2$ theory
\be
\TX  \in \text{SQFT}_{3+1}^{\,\cN=2}
\ee
coupled to a four-manifold $\cM_4$. In what follows we will denote such partition function $Z_{\TX}(\cM_4)$.  Notice that $\TX$ is not necessarily conformal: for instance, if one considers the IIB geometry
\be\label{eq:SYMs}
\CY \equiv \Big\{e^z + e^{-z}  + p_{\fg}(x,y) + u^2 = 0\Big\} \, \subset \mathbb C^4\,,
\ee
the corresponding $\TX$ is the $\cN=2$ supersymmetric Yang-Mills (SYM) theory with simple simply-laced gauge algebra $\fg$. The partition function $Z_{\TX}(\cM_4)$  
should be fully determined by the data of the string background. This is indeed true, but in the case that $\TX$ has non-trivial higher 
symmetries there are important subtleties.

\medskip
The higher $n$-form symmetries in the four dimensional theory act on
the $n$-dimensional defects of the theory. In geometric engineering
these defects arise from D$p$-branes wrapping $k$-cycles on $\CY$,
where $n = p-k+1 \leq 4$. More concretely, any background for the
higher form symmetries introduces monodromies for these defects, so in
the string theory construction such a background must be realized with
a choice of background fluxes. The latter are then providing the
stringy realization of higher form symmetries.\footnote{   This part of
  the discussion, of course, naturally extends to any geometric
  engineering setting. Other recent applications of this formalism to
  different dimensions and amounts of supersymmetry are
  \cite{Garcia-Etxebarria:2019cnb,Morrison:2020ool,Albertini:2020mdx}.}
Therefore, in order to fully specify the theory $\TX$, we need to
specify the fluxes on the IIB construction.

\medskip

It is natural to ask at this point whether there is any canonical
choice for the RR background fluxes. For instance, if we could just
set the flux to zero this would give a canonical choice of $\TX$ for
each geometry $\CY$. Since $\CY$ is a local CY, we must in particular
be aware of the behavior of the flux at infinity.  It turns out that
generically no canonical choice of ``zero flux'' at infinity exists,
rather we have an obstruction which can be argued for as follows. For
any field theory in $d+1$ spacetime dimensions, to each codimension
one submanifold $\cN_{d}$ of spacetime we assign a Hilbert space
$\cH(\cN_{d})$, the space of possible field configurations that are
allowed in the quantum theory on that spatial slice. We obtain this
Hilbert space by quantizing the Hamiltonian of the theory on
$\bR\times \cN_{d}$, where we treat the first coordinate as the time
direction. Similarly, when we consider a string compactification on a
manifold with boundary, we obtain a Hilbert space. In order to
understand which field configurations can be put at infinity, we need
to understand the Hilbert space that type IIB string theory associates
to the boundary $\cN_9\df \cM_4\times Y_5$. Luckily, for our purposes
a rather coarse understanding of this Hilbert space will be enough,
and in particular we will only need to understand its grading by
fluxes.

Our Calabi-Yau threefold $\CY$ is a cone over some Sasaki-Einstein
space $Y_5$, so the choice of boundary conditions at infinity can be
understood as the limiting behaviour of a family of field
configurations at constant radius $r$ in the cone, as
$r\to\infty$. The topology of each slice is $\cM_4\times Y_5$, so the
problem is to understand which choices of field configurations can we
put on $\cM_4\times Y_5$ slices of spacetime.

The relevant grading of Hilbert space was understood by Freed, Moore
and Segal in \cite{Freed:2006yc,Freed:2006ya}. They showed that
whenever $\cN_9$ has torsion,\footnote{  For simplicity we will be
  assuming that $\cM_4$ has no torsion, so the requirement in the text
  is equivalent to $Y_5$ having torsion.} the operators measuring flux
expectation values become non-commutative. More precisely, the
operators measuring electric flux do not commute with the operators
measuring magnetic flux. The $F_5$ flux is self-dual, so in this case
the flux operators for $F_5$ do not commute among themselves. The
precise commutation relation for $F_5$ fluxes is as follows. To each
cohomology class\footnote{  \label{fn:K-theory}More
  carefully, RR fluxes in IIB string theory are believed to be
  classified by elements of the K-theory group $K^1(\cN_9)$
  \cite{Moore:1999gb}. In appendix~\ref{app:K-theory} we show that for
  the geometries that we consider here
  \[
    K^1(\cN_9) = \sum_{n} H^{2n+1}(\cN_9)
  \]
  so in the reminder of the paper we will work in the cohomology
  formulation for simplicity. Unless otherwise specified, all of our
  (co)homology groups are with integer coefficients, so
  $H^n(X)\df H^n(X;\bZ)$.} $\sigma\in \Tor H^5(\cN_9)$ we associate an
operator $\Phi_\sigma$, which measures the torsional part of the flux
on a four-cycle in $\cN_9$ Poincaré dual to $\sigma$. In the usual
(but somewhat imprecise in the current context) differential form
language, these operators would be of the form
$\Phi_\sigma = \exp(i\int_{\cN_9}A_\sigma\wedge F_5)$ where $A_\sigma$
is a flat connection for the torsional class $\sigma$.

The commutation relations between these operators are then
\begin{equation}
  \Phi_\sigma\Phi_{\sigma'} = e^{2\pi i\, \sL(\sigma,\sigma')} \Phi_{\sigma'}\Phi_\sigma
\end{equation}
with
$\sL(\sigma,\sigma')\colon \Tor H^5(\cN_9)\times \Tor H^5(\cN_9) \to
\bQ/\bZ$ the linking number between the cohomology classes (we will
discuss this linking number in more detail below). We refer the reader
to the original papers \cite{Freed:2006yc,Freed:2006ya} for a
derivation of these commutation relations. At any rate, what this
implies for the grading of Hilbert space $\cH(\cN_9)$ by fluxes, and
thus for the available choices of boundary conditions, is that there
is no zero flux eigenstate $\ket{\bf 0}$ such that
$\Phi_\sigma\ket{\bf 0}=\ket{\bf 0}$ for all
$\sigma\in \Tor(H^5(\cN_9))$.

The best that we can do when choosing boundary conditions is to choose
a maximally commuting set $L$ of operators, and impose that our
boundary state is neutral under these. In detail, we define
$s(\sigma,\sigma')\df e^{2\pi i\, \sL(\sigma,\sigma')}$, and define a
maximal isotropic subgroup $L\subset \Tor H^5(\cN_9)$ to be a maximal
set such that $s(\sigma,\sigma')=1$ for all $\sigma,\sigma'\in
L$. This implies that the subgroup generated by the operators
$\{\Phi_\sigma|\sigma\in L\}$ is abelian, and provides a maximal set
of commuting observables. Once we choose $L$, there is a unique state
$\ket{{\bf 0};L}$ in the Hilbert space $\cH(\cN_9)$ such that
$\Phi_\sigma\ket{{\bf 0};L}=\ket{{\bf 0};L}$ for all $\sigma\in L$. We can
interpret this state as follows. Define
\begin{equation}
  F_L\df \frac{\Tor H^5(\cN_9)}{L}
\end{equation}
and choose a representative $\mathbf{f}$ of each coset. Then the
states $\ket{\mathbf{f};L}\df \Phi_{\mathbf{f}}\ket{\mathbf{0};L}$ are
eigenvectors of the flux operators in $L$:
\begin{equation}
  \Phi_\sigma \ket{\mathbf{f};L} = s(\sigma, \mathbf{f})\ket{\mathbf{f};L}\, ,
\end{equation}
so we find that the $\ket{\mathbf{f};L}$ are the states with definite
flux in $L$. In particular, $\ket{\mathbf{0};L}$ can be interpreted as
a state with zero flux in $L$. Once $L$ is chosen this state is
unique, so the non-canonical nature of the choice of boundary
condition reduces to the absence of a canonical choice for $L$ in
$H^5(\cN_9)$.

In this paper we focus on the case $\cN_9=\cM_4\times Y_5$, with
$\Tor H^\bullet(\cM_4)=0$ so by the Künneth formula we have
\begin{equation}
  \Tor H^5(\cN_9) = \bigoplus_{n=0}^4 H^n(\cM_4)\otimes \Tor H^{5-n}(Y_5)\, .
\end{equation}
In the cases of interest to us we additionally have that $Y_5$ is
simply connected \cite{10.2307/j.ctt1bd6kvv}, so
$\Tor H_1(Y_5)=\Tor H^4(Y_5)=0$. The universal coefficient theorem
\cite{Hatcher:478079} additionally implies that
$\Tor H^2(Y_5)=\Tor H_1(Y_5)=0$, so the only possible non-trivial
torsion lives in $H^3(Y_5)$:
\begin{equation}
  \Tor H^5(\cN_9) = H^2(\cM_4)\otimes \Tor H^3(Y_5) \cong H^2(\cM_4; \Tor H^3(Y_5))\, .
\end{equation}

In principle we should now classify all the $L\subset \Tor H^5(\cN_9)$
for every $\cN_9$, but there is a class of such isotropic subgroups
which is particularly interesting in the context of four dimensional
physics on $\cM_4$. Assume that we fix a $Y_5$, or equivalently its
cone $X_6$. Then there is a subclass of the possible $L$ of the form
\begin{equation}
  \label{eq:genuine-polarizations}
  L = L_5\otimes H^2(\cM_4)
\end{equation}
that can be defined uniformly for every $\cM_4$.\footnote{  At least for
  those $\cM_4$ without torsion. We do not know if a similar canonical
  subset of choices exists if we allow for torsion in $\cM_4$.} Here
$L_5$ is a maximal isotropic subgroup of $\Tor H^3(Y_5)$. The theories
defined by such choices are sometimes called ``genuine'' four
dimensional theories.\footnote{  Choices of boundary conditions outside
  this class will depend on specific features of $\cM_4$. This is
  perfectly fine from the IIB point of view, but the four dimensional
  interpretation of the resulting theories is slightly less conventional. We
  refer the reader to \cite{Garcia-Etxebarria:2019cnb} for a more
  detailed discussion of this point.}

The choices of global structure for the genuine $\TX$ theories are
thus the choices of maximal isotropic $L_5\subset \Tor H^3(Y_5)$, with
$\partial \CY=Y_5$. Once we have such an $L_5$ we have a choice for
the 2-surface operators generating the 1-form symmetries of $\TX$:
they come from the reduction of the $\Phi_\sigma$ flux operators in
the IIB theory. And relatedly, introducing background fluxes for $F_5$
at infinity will introduce background fluxes for the 1-form symmetries
in the four-dimensional theory on $\cM_4$.

\subsection{The case of hypersurface singularities}

Our discussion so far has been fairly general, and has not required us
to make use of the fact that the $\TX$ theories preserve $\cN=2$. In
fact, in addition to being $\cN=2$ supersymmetric the theories that we
will be discussing have the nice property that their BPS spectrum can
be generated from a BPS quiver \cite{Cecotti:2011rv, Alim:2011kw} (we
refer the reader unfamiliar with BPS quivers to these papers for
reviews), and this leads to a reformulation of the answer that we just
found in terms of screening of line operators, generalizing to our
current context the discussions in
\cite{Aharony:2013hda,DelZotto:2015isa}.

Recall that each node in the BPS quiver represents a BPS building
block, and the arrows encode how they can be recombined. From the IIB
perspective, the nodes in the quiver represent D3 branes wrapped on
generators of a basis of $H_3(\CY)$, and the arrows in the quiver
encode the intersection numbers of the corresponding 3-cycles. BPS
states in $\TX$ can be obtained from D3 branes wrapping
supersymmetric compact cycles in $\CY$, and such D3 branes can always
be constructed by taking a combination of generators with the right
total charge, and recombining them.\footnote{  More formally, we have
  that the derived category of A-branes on $X_6$ is isomorphic to
  the derived category of representations of the quiver. We refer the
  reader to \cite{Aspinwall:2004jr} for a review of this
  approach.\label{fn:brane-category}}

We can connect our discussion in the previous section to the
formulation in terms of BPS quivers as follows. Take a small
7-dimensional sphere $S^7$ around the origin in $\bC^4$. Our
``boundary at infinity'' $Y_5$ is homotopy equivalent to the
intersection $\cY_5=S^7\cap\{P_{(\fg,\fg')}=\epsilon\}$, where
$P_{(\fg,\fg')}$ was the
polynomial~\eqref{eq:Argyres-Douglas-geometry} defining the
Argyres-Douglas theory, and we take $\epsilon$ small but non-vanishing
in order to make the interior of $\cY_5$ smooth. Finally, introduce
$\cX_6$ to be the (smooth) interior of $\cY_5$, namely the
intersection of a ball $B^8$ (such that $\partial B^8=S^7$) with
$P_{(\fg,\fg')}=\epsilon$.

Since $\partial \cX_6=\cY_5$ there is a long exact sequence in
homology of the form
\begin{equation}
  \ldots \to H_n(\cY_5) \to H_n(\cX_6) \to H_n(\cX_6,\cY_5) \to H_{n-1}(\cY_5) \to \ldots
\end{equation}
where $H_n(\cX_6,\cY_5)$ denotes relative homology, and the maps in the
same degree are the obvious ones.

We are interested in $\Tor H^3(\cY_5)=\Tor H_2(\cY_5)$. It
is a classical result of Milnor (see theorems 5.11 and 6.5 of
\cite{10.2307/j.ctt1bd6kvv}) that $\cX_6$ has the homotopy type of a
bouquet of three-spheres, and in particular $H_2(\cX_6)=0$. The long
exact sequence in homology above then implies that
\begin{equation}
  \label{eq:defect-group}
  H_2(\cY_5) = \frac{H_3(\cX_6,\cY_5)}{H_3(\cX_6)}
\end{equation}
where the embedding of $H_3(\cX_6)$ into $H_3(\cX_6,Y_5)$ is the
natural one. This equation has a natural interpretation in terms of
four dimensional field theory, as follows. The numerator denotes the
homology class of 3-cycles in $\cX_6$, including those that extend to
the boundary. If we wrap D3 branes on these cycles we obtain lines in
the four dimensional theory. The D3 branes wrapping compact 3-cycles
in $\cX_6$ give dynamical lines, while the ones extending to the
boundary give line defects. The denominator includes the dynamical
lines only, so~\eqref{eq:defect-group} is saying that in order to
understand the global structure of the theory, we need to consider the
line defects modulo the dynamical excitations, or in other words the
unscreened part of the line defect charge, as in
\cite{Aharony:2013hda,DelZotto:2015isa}.

It is convenient to rephrase the previous discussion in the language
of cohomology groups. Lefschetz duality implies that
$H_3(\cX_6,\cY_5)\cong H^3(\cX_6)$, and the universal coefficient
theorem then implies that
$H_3(\cX_6,\cY_5)\cong\Hom(H_3(\cX_6),\bZ)$. On the other hand, the
embedding of $H_3(\cX_6)$ into $\Hom(H_3(\cX_6),\bZ)$ is given simply by
the partial evaluation of the intersection form
$q\colon H_3(\cX_6)\times H_3(\cX_6)\to \bZ$. That is, given any element
$x\in H_3(\cX_6)$ we have an embedding
$Q\colon H_3(\cX_6)\to \Hom(H_3(\cX_6),\bZ)$ given by $Q(x) =
q(x,\cdot)$. We can thus rewrite~\eqref{eq:defect-group} as
$H_2(\cY_5) = \coker(q)$, or equivalently
\begin{equation}
  \label{eq:homology-coker}
  H_2(\cY_5)=\coker(Q)\, .
\end{equation}

Now, $Q$ is an integer-valued antisymmetric matrix, so there is a
change of basis to $\sQ$ (that is, an integer matrix $P$ with
$\det(P)=\pm 1$ such that $Q=P^t \sQ P$) with
(see theorem IV.1 in \cite{newman1972integral})
\begin{equation}
  \label{eq:canonical-Q}
  \sQ = \begin{pmatrix}
    0 & r_1 \\
    -r_1 & 0\\
    && 0 & r_2\\
    && -r_2 & 0 \\
    &&&& \ddots\\
    &&&&& 0 & r_n\\
    &&&&& - r_n & 0\\
    &&&&&&& 0\quad \,\\
    &&&&&&&& \ddots \\
    &&&&&&&&& \quad 0
  \end{pmatrix}
\end{equation}
and $r_i \in \bZ$, such that $r_i\mid r_{i+1}$. Without loss of
generality we can choose $r_i> 0$. Since $P$ is invertible we have
$\coker(Q)=\coker(\sQ)$.  Let us focus on a single $2\times 2$ block
in $\sQ$ of the form
\begin{equation}
  \sQ_i = \begin{pmatrix}
    0 & r_i\\
    -r_i & 0
  \end{pmatrix}
\end{equation}
with $r_i>0$. We denote the generators of $H_3(\cX_6)$ on this
subspace $a,b$, and the dual elements $a^*,b^*\in
\Hom(H_3(\cX_6))$. We have $\sQ_i(a)=r_i b^*$ and $\sQ_i(b)=-r_i
a^*$. This implies that $\coker(\sQ_i)=\bZ_{r_i}\oplus \bZ_{r_i}$. We
thus have
\begin{equation}
  \coker(\sQ)=\bZ^\kappa\oplus \sum_{i=1}^n \bZ_{r_i} \oplus \bZ_{r_i}
\end{equation}
with $$\kappa=\rk(H_3(\cX_6))-2n = \rk F$$
is the factor corresponding to the 0-form flavor symmetry of the theory, while $n$ is its rank (i.e. the dimension of the Coulomb branch of the SCFT). This determines $\Tor H_2(\cY_5)$
as an abelian group
\begin{equation}
  \Tor H_2(\cY_5) = \sum_{r_i>1} \bZ_{r_i} \oplus \bZ_{r_i}\, .
\end{equation}
The case $r_i=1$ is trivial; we choose to exclude it from the sum. 

In order to understand the global structure of the Argyres-Douglas
theories we need a final piece of additional information, the linking
pairing between elements in $H_2(\cY_5)$. Let us focus again on a
single block $\sQ_i$, with $r_i > 1$. The linking form $\sL_i$ on
$[\Tor H_2(\cY_5)]_i$ (that is, the $i$-th block of $\Tor H_2(\cY_5)$) is related very simply to $\sQ_i$ \cite{GordonLitherland}:
\begin{equation}
  \label{eq:Qi-pairing}
  \sL_i = \sQ_i^{-1} = \begin{pmatrix} 0 & -\frac{1}{r_i}\\
    \frac{1}{r_i} & 0 \end{pmatrix} \mod 1\, .
\end{equation}

The final answer from our analysis is thus quite
straightforward. Recall from~\eqref{eq:genuine-polarizations} that we
are after maximal isotropic subgroups of $\Tor H_2(\cY_5)$, where the
commutation relations are determined by the linking form $\sL$. From
the form above, the problem then reduces to the classification of the
maximal isotropic sublattices for each block of $\sQ_i$, that is the
maximal isotropic sublattices of $\bZ_{r_i}\oplus \bZ_{r_i}$ with the
pairing~\eqref{eq:Qi-pairing}. (This problem is isomorphic to the
problem of determining the global forms of the $\cN=4$ $\fsu(r_i)$
theory, studied in \cite{Aharony:2013hda}.)

As an example, assume that
\begin{equation}
  \sQ = \begin{pmatrix}
    0 & 2\\
    -2 & 0\\
    && 0 & 2\\
    && -2 & 0
  \end{pmatrix}\, .
\end{equation}
As we will show below, $\cT[A_4,D_6]$ is of this type. Each
$2\times 2$ block is of the form
\begin{equation}
  \sQ_i = \begin{pmatrix}
    0 & 2\\
    -2 & 0
  \end{pmatrix}
\end{equation}
leading to a contribution to the torsion of the form
$\bZ_2\oplus \bZ_2$, so in total $\Tor H_2(\cY_5)=\bZ_2^4$. For each
block, we have three maximal isotropic subgroups, which in this case
comprise a single element. They are $\{(1,0)\}$, $\{(0,1)\}$ and
$\{(1,1)\}$. (At this point we encourage the reader to compare with
the global forms of the $\fsu(2)$ theory in \cite{Aharony:2013hda}.)
So, we find that there are in total $3\times 3 = 9$ possible choices
for the global form of the $\cT[A_4,D_6]$ theory.

\subsection{A categorical aside}\label{sec:sgroto}

There is an interesting question to be asked at this point: we are
discussing screening of branes by branes, but we are working at the
level of cohomology classes, while a full description of branes
requires more information than this. For instance, as in
footnote~\ref{fn:brane-category}, we may want to think of branes as
elements of a suitable derived category. How do we implement the
notion of screening in this categorical language? A beautiful answer
to this question has been provided in \cite{Caorsi:2017bnp}, and it
seems to agree nicely with our results: the torsional part of the
group of screened branes is given by the torsional part of the
Grothendieck group $K_0(\cC(\Gamma))$ (known as the ``'t Hooft group''
in that paper), where $\Gamma$ is the Ginzburg algebra of the quiver
with superpotential at hand, and $\cC(\Gamma)$ its associated
cluster algebra. There is a short exact sequence
\begin{equation}
  0\to D^b\Gamma \to \mathfrak{Per}\,\Gamma \to \cC(\Gamma)\to 0
\end{equation}
in which $D^b\Gamma$ is the derived category describing (subject to an
additional stability condition) the BPS spectrum of the theory, while
$\mathfrak{Per}\,\Gamma$ denotes the category of all line
defects. This is all very reminiscent of our geometric discussion, so
whenever we have a geometric engineering for the theory, we expect the
categories $D^b\Gamma$ and $\mathfrak{Per}\,\Gamma$ to refine the
purely homological information about the branes in $H_3(\cX_6)$ and
$H_3(\cX_6,\cY_5)$, respectively. We will not attempt to put this
correspondence on a firmer footing in this paper, but let us point out
that the key implication of the correspondence for our purposes,
namely that the Grothendieck group $K_0(\cC(\Gamma))$ equals the
internal part of the group of fluxes at infinity
$K_0(\cY_5)=K^1(\cY_5)=\coker(Q)$, is actually true in the case of
cluster categories associated to ``hereditary'' categories
\cite{BAROT200833}. Moreover, the Weyl pairing introduced in \S 4.3.6 of  \cite{Caorsi:2017bnp} coincides with the linking pairing we find here. 
 We refer the reader to that paper for the
definition of this class of theories, and to \S2.6.2 of
\cite{Caorsi:2017bnp} for specific examples of physically interesting
theories of this kind.

\medskip

At this point we can simply remark that the result of \cite{Caorsi:2017bnp} about the 't Hooft group from the BPS quivers $Q$ associated to 4d $\cN=2$ SYM, interpreted as results about the defect groups for the geometry in \eqref{eq:SYMs} as discussed above are perfectly consistent with the possible global structures of pure $\cN=2$ SYM from \cite{Aharony:2013hda}. This serves as a further nice  consistency check for our formalism, that goes beyond the case of hypersurface singularities in $\bC^4$.

\section{Global structure of the $(\fg,\fg')$ Argyres-Douglas theories}\label{sec:GGPRIME}

We now want to apply the ideas developed in the previous sections to
the particular case of the Argyres-Douglas theories. The BPS quivers
for these theories are known \cite{Cecotti:2010fi, Keller2010THEPC},
so in principle we already have all the information that we need at
hand, but in order to give general results it is more convenient to
use results by Boyer, Galicki and Simanca \cite{Boyer} that we now
summarize. An additional benefit is that these results also apply to
some examples beyond the $(\fg,\fg')$ theories that will be of
interest below, and whose BPS quiver has not appeared previously in
the literature.

\subsection{The torsion at infinity for quasi-homogeneous threefold
  singularities}

Let $Y_5$ be, as above, a 5d manifold homotopic to the boundary at
infinity. We model it by $f^{-1}(0)\cap S^7$, namely the intersection
of the hypersurface $V=\{x=(x_1,x_2,x_3,x_4)\;|\;f(x)=0\}$ and the
$7$-sphere $S^7$ inside $\bC^4$, where $f=f(x)$ is a quasi-smooth
\cite{fletcher} weighted homogeneous polynomial with weights
$w=(w_1,w_2,w_3,w_4)$ and of total degree $d$, with isolated
singularities at the origin. All of the examples that we will discuss
in this paper are of this type. We have that \cite{OM,Boyer}
\begin{equation}
  \label{eq:BGS}
  H_2(Y_5)= \mathbb Z^\kappa \oplus \sum_{i=1}^4\mathbb{Z}_{r_i}^{2g_i}\, ,
\end{equation}
where
\be\label{eq:kappa1}
\kappa = \sum (-1)^{4-s} {u_{i_1} \cdots u_{i_s} \over v_{i_1} \cdots v_{i_s} \, \text{lcm}(u_{i_1}\cdots u_{i_s})}  \qquad
u_i = {d \over \text{gcd}(d,w_i)} \qquad v_i = {w_i \over \text{gcd}(d,w_i)} 
\ee
and the sum is taken over all the 16 subsets $\{i_1,...,i_s\}$ with $s$ elements of the index set $\{1,2,3,4\}$. Moreover,
\be
r_i=\text{gcd}(w_1,..,\hat{w}_i,...,w_4)
\ee
and
\be
2g_i=-1+\sum_{j\neq i}\frac{\text{gcd}(d,w_j)}{w_j}-
d\sum_{j< k\, ,\, j,k\neq i}\frac{\text{gcd}(w_j,w_k)}{w_jw_k}+d^2 \frac{r_i}{w_1...\hat{w}_i...w_4}\, ,
\end{equation}
and the notation $\hat{w}$ means omit $w$. We have summarized the
results of applying these formulas to the $(\fg,\fg')$ Argyres-Douglas
theories in table~\ref{table:Argyres-Douglas}. 

\medskip

In general $\kappa$ can be determined from worldsheet/target correspondence \cite{Cecotti:2010fi} using the method discussed in \cite{Cecotti:2011gu} as was done in \cite{DelZotto:2014kka} for the case of $\mathcal{T}[\mathfrak a_n, \mathfrak a_m]$: $\kappa$ corresponds to the number of +1 eigenvalues of the 2d $tt^*$ monodromy for the associated 2d (2,2) LG model. The latter coincides with the monodromy associated to the Milnor fibration of the singularity, and hence can be computed geometrically: $\kappa$ is just the number of $\Phi_1(Z) = (Z-1)$ factors in the Alexander polynomial of the singularity, which is nothing but the characteristic polynomial of the 2d $tt^*$ monodromy \cite{Cecotti:1992rm}. Now for $\Tgg$ theories, we have that the 2d $tt^*$ monodromy is simply given by $H_\fg \otimes H_{\fg'}$, where the latter are the monodromies corresponding to the minimal $\fg$ and $\fg'$ singularities \cite{Cecotti:1992rm}. $H_\fg$ has eigenvalues 
\be
\exp 2 \pi i \Big( {\ell_i(\fg) \over h^\vee(\fg)} - {1\over 2}\Big)  \qquad i = 1,...,r_\fg
\ee
where $h^\vee(\fg)$ is the dual Coxeter number of $\fg$, $r_\fg$ its rank, and $\ell_i(\fg)$ are its exponents. Therefore $H_\fg \otimes H_{\fg'}$ has eigenvalues
\be
\exp 2 \pi i \Big( {\ell_i(\fg) \over h^\vee(\fg)} +  {\ell_j(\fg') \over h^\vee(\fg')}\Big)\,.
\ee
This gives a simple formula for the rank of the flavor symmetry of $\Tgg$:
\be\label{eq:kappa2}
\kappa = \# \left\{ \text{solutions of }  {\ell_i(\fg) \over h^\vee(\fg)} +  {\ell_j(\fg') \over h^\vee(\fg')} \in \mathbb Z \atop \text{as } i=1,...,r_\fg \text{ and } j=1,..,r_{\fg'}\right\}\,.
\ee
For the case of $\mathcal{T}[\mathfrak a_n, \mathfrak a_m]$ this formula can be easily evaluated \cite{DelZotto:2014kka}:
\be
\kappa_{\,\mathcal{T}[\mathfrak a_n, \mathfrak a_m]} = \text{gcd}(h^\vee(\mathfrak a_n), h^\vee(\mathfrak a_m))-1\,.
\ee
This expression must coincide with the expression for $\kappa$ obtained from singularity theory, because that is defined as the order of $(X-1)$ in the Alexander polynomial of the singularity, hence all seemingly different expressions for $\kappa$, namely equations \eqref{eq:kappa1}, \eqref{eq:kappa2}, and $\rk \coker \sQ$  have to agree.

\begin{table}
\centering

\begin{tabular}{  m{2.5cm} |  m{3cm} m{1.6cm} m{0.5cm} m{3cm}  }\hline

$(g,g')$ & $ \Tor H_2(Y_{(\fg,\fg')})$ \\ [0.1cm]
\hline\hline
$(A_n,A_m)$ & 0 \\ [0.1cm] 
\hline
\multirow{3}{*}{$(A_{n-1},D_{m+1})$} 
         &$\mathbb{Z}_2^{\gcd(n,m)-1}$ & $\text{if }  \phantom{m}2 \nmid n$\\
         & $\mathbb{Z}_2^{\gcd(n,m)-2}$ & $\text{if }  \phantom{m}2 \mid m$ & $\text{and}$ & $\gcd(n,2m) \mid m$\\
         &0 & otherwise\\
\hline
\multirow{4}{*}{$(A_{n-1},E_6)$}
    &0 & $\text{if } 12 \mid n$\\
    &$\mathbb{Z}_2^2$ & $\text{if }  \phantom{1}6 \mid n$\\
    &$\mathbb{Z}_3^2$ & $\text{if }  \phantom{1}4 \mid n$\\
    &0&$\text{otherwise}$\\
\hline
\multirow{4}{*}{$(A_{n-1},E_7)$}
    &0 & $\text{if } 18 \mid n$\\
    &$\mathbb{Z}_2^6$ & $\text{if }  \phantom{1}9 \mid n$\\
    &$\mathbb{Z}_3^2$ & $\text{if }  \phantom{1}6  \mid n$\\
    &0&$\text{otherwise}$\\ 
\hline
\multirow{5}{*}{$(A_{n-1},E_8)$}
    &0 & $\text{if } 30 \mid n$\\
    &$\mathbb{Z}_2^8$ & $\text{if } 15  \mid n$\\
    &$\mathbb{Z}_3^4$ & $\text{if } 10  \mid n$\\
    &$\mathbb{Z}_5^2$ & $\text{if } \phantom{1}6  \mid n$\\
    &0&$\text{otherwise}$\\
\hline
\multirow{5}{*}{$(D_{n+1},D_{m+1})$}
    &$\mathbb{Z}_2^{\gcd(n,m)}$ &
    $\text{if } \phantom{1}2 \mid m$ &
    $\text{and}$ & $(2\gcd(n,m)) \mid n$\\
    &$\mathbb{Z}_2^{\gcd(n,m)}$ &
    $\text{if } \phantom{1}2 \mid n$ &
    $\text{and}$ &
    $(2\gcd(n,m)) \mid m$ \\
    &$\mathbb{Z}_2^{\gcd(n,m)-1}$ &
    $\text{if } \phantom{1}2 \mid n$ &
    $\text{and}$ &
    $\phantom{1}2 \nmid m$ \\
    &$\mathbb{Z}_2^{\gcd(n,m)-1}$
    & $\text{if } \phantom{1}2 \nmid n$
    & $\text{and}$
    & $\phantom{1}2 \mid m$\\
    & 0 & $\text{otherwise}$\\
\hline
\multirow{5}{*}{$(D_{n+1},E_6)$}
    &$\mathbb{Z}_2^6$ & $\text{if } 12 \mid n$\\
    &0 & $\text{if } \phantom{1}4 \mid n$ & $\text{or}$ &  $\phantom{1}6 \mid n$\\
    &$\mathbb{Z}_4^2$ & $\text{if } \phantom{1}3 \mid n$ \\    &$\mathbb{Z}_3^2$ & $\text{if } \phantom{1}2 \mid n$ \\
    &0&$\text{otherwise}$\\
\hline
\multirow{5}{*}{$(D_{n+1},E_7)$}
    &$\mathbb{Z}_2^6$ & $\text{if } 18 \mid n$\\
    &0 & $\text{if}\; \phantom{1}9 \mid n$ & $\text{or}$ &  $\phantom{1}6 \mid n$\\
    &$\mathbb{Z}_3^2$ & $\text{if } \phantom{1}3 \mid n$\\
    &0&$\text{otherwise}$\\
\hline
\multirow{5}{*}{$(D_{n+1},E_8)$}
    &$\mathbb{Z}_2^8$ & $\text{if } 30 \mid n$\\
    &0 & $\text{if}\; 15 \mid n$ & $\text{or}$ & $10 \mid n \quad \text{or}\; \phantom{1}6 \mid n$\\
    &$\mathbb{Z}_3^4$ & $\text{if } \phantom{1}5 \mid n$\\
    &$\mathbb{Z}_5^2$ & $\text{if } \phantom{1}3 \mid n$\\
    &0&$\text{otherwise}$\\
\hline
\multirow{1}{*}{$(E_n,E_m)$}
    &0\\
\end{tabular}%
\caption{Defect groups for the Argyres-Douglas theories. Whenever two
  cases overlap the earliest applicable one is the correct result. For instance,
  $\Tor(H_2(Y_{A_{11},E_6}))=0$.}
\label{table:Argyres-Douglas}
\end{table}
\clearpage

\subsection{On $\cN=1$ Lagrangians for $\cN=2$ theories}
\label{sec:Maruyoshi-Song}

As an application of the results in the previous section, we will now
check that the $\cN=1$ UV Lagrangians for various $(\fg,\fg')$
Argyres-Douglas theories proposed by Agarwal, Sciarappa and Song
\cite{Agarwal:2017roi} (building on work by Maruyoshi and Song
\cite{Maruyoshi:2016aim,Maruyoshi:2016tqk}; see also
\cite{Giacomelli:2017ckh, Carta:2020plx} for recent closely related work)
have global structure compatible with the results we just found.  More
precisely, we will be checking that the set of choices of global
structure for the IR theory is in one-to-one correspondence with the
choices for the UV theory.

As an example, consider the $\cN=2$ theory
\begin{equation}
  \boxed{\fso(2)}-\fsp(2N)-\fso(4N+2)-\fsp(6N)- \ldots -\fsp(4pN-2N)-\boxed{\fso(4pN+2)}
\end{equation}
which was argued in \cite{Agarwal:2017roi} to flow, upon a suitable
$\cN=1$ mass deformation, to the $(A_{2p-1}, D_{2Np+1})$
Argyres-Douglas theory. We will denote this Lagrangian theory
$\cL[A_{2p-1}, D_{2Np+1}]$. The boxed factors at the ends indicate
global symmetry factors, and the lines between adjacent $\fso$ and
$\fsp$ nodes indicates a hypermultiplet in the vector representation
of $\fso$ and in the fundamental of $\fsp$. There are in total $p$
nodes carrying a $\fsp$ gauge factor, and $p-1$ nodes carrying a
$\fso$ gauge factor.

Our task is to classify all the possible global forms of the $\cN=1$
theory arising from the mass deformation of this theory. Note that the
$\cN=1$ mass deformation is introduced by giving a vev to local
operators in the $\cN=2$ Lagrangian, so the choices of global
structure of the $\cN=1$ theory are identical to those of the $\cN=2$
theory without the mass deformation.

To classify the global forms of $\cL[A_{2p-1}, D_{2Np+1}]$ we can
proceed as in \cite{Aharony:2013hda}. First, denote by $G$ the
universal cover of the gauge group of $\cL[A_{2p-1}, D_{2Np+1}]$:
\begin{equation}
  G = \Sp(2N)\times Spin(4N+2)\times \cdots \times \Sp(4pN - 2N)\, .
\end{equation}
Note that we do not include flavour symmetries here. Each $\Sp$
factor has a $Z(\Sp(2k))=\bZ_2$ center, and each $Spin(4k+2)$ factor
a $Z(Spin(4k+2))=\bZ_4$ center, so in all:
\begin{equation}
  Z(G) = \bZ_2 \oplus \bZ_4 \oplus \cdots \oplus \bZ_2\, .
\end{equation}
Denote by $Z^0$ the subgroup of the center that leaves all matter
hypermutiplets invariant. The vector representation of $Spin(4k+2)$
has charge 2 under the generator of $Z(Spin(4k+2))=\bZ_4$ (the spinor
representation has charge 1), so a $\bZ_2$ subgroup of each
$Z(Spin(4k+2))=\bZ_4$ leaves matter invariant. Together with the fact
that we have matter transforming in the fundamental of $\fsp(2N)$ and
$\fsp(4pN-2N)$ at the ends of the quiver this implies that we will
have a $\bZ_2$ contribution to $Z^0$ from each $Spin$ factor of $G$,
and nothing from the $\Sp$ factors. We have $(p-1)$ $Spin(4k+2)$
factors in $\cL[A_{2p-1}, D_{2Np+1}]$, so $Z^0=\bZ_2^ {p-1}$.

We can now apply the prescription of \cite{Aharony:2013hda}
straightforwardly. Take two copies of $Z^0$, which we denote
$Z^0_{\text{electric}}$ and $Z^0_{\text{magnetic}}$. Both are equal to
$\bZ_2^{p-1}$ in our case. Construct the lattice
\begin{equation}
  L_{\cL} = Z^0_{\text{electric}} \oplus Z^0_{\text{magnetic}} = \bZ_2^{p-1} \oplus \bZ_2^{p-1}
\end{equation}
of (screened) line operator charges. The pairing between such charges
can be put in block diagonal form, with each block involving a
$\bZ_2\oplus \bZ_2$ coming from Wilson and 't Hooft lines charged
under the $\bZ_2$ contributed by each $Spin(4k+2)$ factor. That is,
the pairing inside each block is of the form
\begin{equation}
  \label{eq:Z2-block-pairing}
  \dsz{\cdot}{\cdot}_{Spin(4k+2)} = \begin{pmatrix} 0 & 1\\
    1 & 0 \end{pmatrix} \mod 2\, .
\end{equation}
The set of global forms will be the set of maximal isotropic subspaces
of $L_{\cL}$ with this pairing.

Now we just need to compare with the result from
table~\ref{table:Argyres-Douglas} for the torsion of the putative IR
fixed point, namely $\cT[A_{2p-1},D_{2Np+1}]$. In the notation of that
table we have $n=2p$ and $m=2Np$, we find that
\begin{equation}
  \Tor H_2(Y_{(A_{2p-1},D_{2Np+1})}) = \bZ_2^{2p-2}\, .
\end{equation}
This perfectly agrees with the result for $L_{\cL}$ for
$\cL[A_{2p-1},D_{2Np+1}]$ found above. Furthermore, the pairing of
fluxes induced by the linking pairing on
$\Tor H_2(Y_{(A_{2p-1},D_{2Np+1})})$ can always be put in the
block diagonal form~\eqref{eq:Z2-block-pairing}
\cite{Garcia-Etxebarria:2019cnb}. So at this point the problem of
classifying maximal isotropic subgroups of $L_{\cL}$ becomes
isomorphic to the problem of classifying maximally commuting flux
operators on the boundary of the IIB construction, and we are done.

Other cases in \cite{Agarwal:2017roi} can be dealt with similarly. For
instance, \cite{Agarwal:2017roi} propose that a suitable mass
deformation of the $\cN=2$ Lagrangian
\begin{equation}
  \fso(2k)-\fsp(4k-4)-\fso(6k-4)-\fsp(8k-8)- \ldots - \fsp(4p(k-1)) - \boxed{\fso(4p(k-1)+2k)}
\end{equation}
flows to $\cT[A_{2p},D_{2p(k-1)+k}]$. In this case there are
in total $p$ gauge nodes carrying a $\fsp$ algebra and $p$ carrying a
$\fso$ algebra. The analysis of the Lagrangian theory proceeds just as
above, with the result that
\begin{equation}
  L_{\cL} = \bZ_2^p \oplus \bZ_2^p
\end{equation}
and block diagonal pairing as in~\eqref{eq:Z2-block-pairing}. (We will
henceforth omit the pairing from our discussion, since it is always of
this block diagonal form.)

We should now compare with the results in
table~\ref{table:Argyres-Douglas}. We find
\begin{equation}
  \Tor H_2(Y_{A_{2p},D_{2p(k-1)+k}}) = \bZ_2^ {\gcd\bigl(2p+1, (2p+1)(k-1)\bigr)-1} = \bZ_2^{2p}
\end{equation}
which again agrees perfectly with $L_{\cL}$.

Within the class of $(\fg,\fg')$ theories, \cite{Agarwal:2017roi} also
predicts that a suitable $\cN=1$ deformation of the $\cN=2$ Lagrangian
theory with quiver
\begin{equation}
  \fsu(N)-\fsu(2N)-\fsu(3N)-\ldots - \fsu((p-1)N) - \boxed{\fsu(pN)}
\end{equation}
flows to $\cT[A_{p-1}, A_{Np-1}]$. In this case it is clear that
$\cL_{\cL}=0$, which agrees with the result for
$\Tor H_2(Y_{(A_{p-1}, A_{Np-1})})$ in table~\ref{table:Argyres-Douglas}.

\medskip

For completeness, let us mention the two cases in
\cite{Agarwal:2017roi} that are not of $(\fg,\fg')$ type. First, the
$(I_{p,pk},S)$ theories have a UV Lagrangian description given by the
quiver
\begin{equation}
  \boxed{\fsu(1)} - \fsu(k+1)-\fsu(2k+1)-\ldots - \fsu((p-1)k+1) - \boxed{\fsu(pk+1)}\, .
\end{equation}
It is easy to see that in this case $L_{\cL}=0$. The IIB geometry
engineering this theory is of the form \cite{Wang:2015mra,Xie:2016evu}
\begin{equation}
  x_1^2 + x_2^2 + x_3^{pk+1} + x_3z^p = 0
\end{equation}
and one can check easily from here that~\eqref{eq:BGS} gives
\begin{equation}
  \Tor Y_{(I_{p,pk},S)} = 0
\end{equation}
in agreement with the prediction of the UV theory.

Finally, perhaps more interestingly, we have the UV Lagrangian
theories with quiver of the form
\begin{equation}
  \fsp(2N) - \fso(4N+4) - \fsp(6N+4) - \fso(8N+8) - \ldots - \fsp((p-1)(2N+2)+N) - \boxed{\fso(4p(N+1))}\, .
\end{equation}
There are $(p-1)$ factors of $\fso$ type in the gauge group, and $p$
factors of $\fsp$ type. By the same reasoning as in previous cases we
find that
\begin{equation}
  L_{\cL} = \bZ_2^{p-1} \oplus \bZ_2^{p-1}\, .
\end{equation}
Upon a suitable $\cN=1$ preserving mass deformation,
\cite{Agarwal:2017roi} find convincing evidence that this theory flows
to the $D_{2p(N+1)}^{(2p(N+1))}[p]$ theories introduced in
\cite{Wang:2015mra}. Slightly more generally, consider the theories
defined by IIB on
\begin{equation}
  x^2 + y^n + yz^2 + zw^k = 0\, .
\end{equation}
The resulting SCFTs were called $D_{n+1}^{(n+1)}[k]$ in
\cite{Wang:2015mra}. The topology at infinity for this geometry has
torsion
\begin{equation}
  \Tor H_2\left(Y_{D_{n+1}^{(n+1)}[k]}\right) = \begin{cases}
    \bZ_2^{\gcd(2k,n+1)-2} & \text{if}\; 2 \mid \frac{n+1}{\gcd(n+1,k)}\\
    0&\text{otherwise}
  \end{cases}
\end{equation}
In particular, we have
\begin{equation}
  \Tor H_2\left(Y_{D_{2p(N+1)}^{(2p(N+1))}[p]}\right) = \bZ_2^{2p-2}
\end{equation}
again reproducing the results from the proposed UV description.

\acknowledgments

This project has received funding from the European Research Council
(ERC) under the European Union's Horizon 2020 research and innovation
programme (grant agreement No. 851931). I.G.-E. is partially supported
by STFC consolidated grant ST/P000371/1. S.S.H. is supported
by the STFC grant ST/T506035/1.

\appendix

\section{K-theory groups for the boundary of isolated threefold singularities}

\label{app:K-theory}

\begin{figure}
  \centering
  \includegraphics{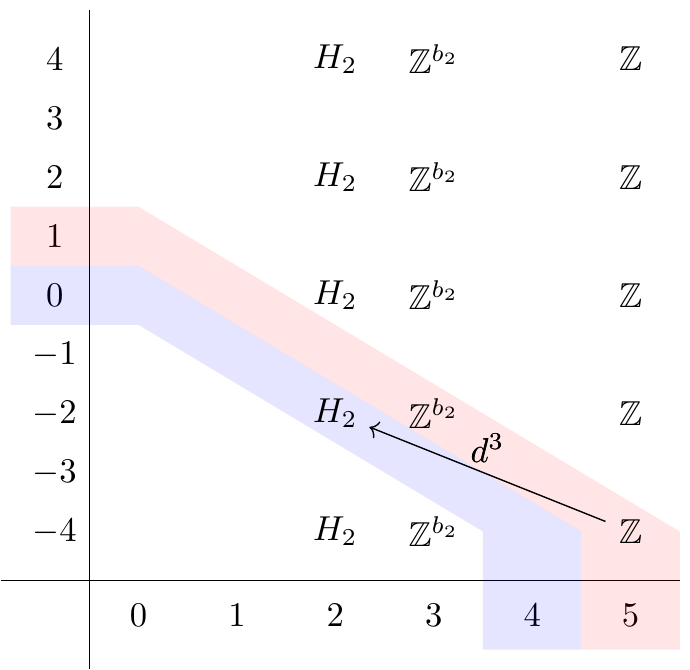}
  \caption{Second page for the Atiyah-Hirzebruch spectral sequence for
    the reduced K-homology of the horizon manifold on an isolated
    hypersurface singularity. We have denoted $H_2\df H_2(Y^5)$,
    $b_2\df \rk (H_2\otimes \bQ)$, and shown the only differential
    that might potentially be non-vanishing. The entries shaded in
    blue are those contributing to $K_0(Y_5)$, and those in pink are
    those contributing to $K_1(Y_5)$.}
  \label{fig:AHSS-K-homology}
\end{figure}

In this appendix we will compute the K-theory groups of
the manifold $Y_5$ at the boundary of an isolated hypersurface
singularity. It is convenient to do so by computing the reduced
Atiyah-Hirzebruch spectral sequence for homology (see remark 2 in
pg. 351 of \cite{Switzer})
\begin{equation}
  E_2^{p,q} = \tilde H_p(Y_5;K_q(\pt)) \Longrightarrow \tilde K_{p+q}(Y_5)\, .
\end{equation}
The second page of this spectral sequence is shown in
figure~\ref{fig:AHSS-K-homology}. Note that in writing that spectral
sequence we are using $H^1(Y_5)=H^4(Y_5)=0$.

The only potentially non-vanishing differential is indicated by $d^3$
in the drawing. This is the first non-vanishing differential, so it is
a stable homology operation (see \S4.L in \cite{Hatcher:478079} for a
definition and proofs of some of the statements below) dual to a
stable cohomology operation $d_3 \colon H^0(Y_5)\to H^3(Y_5)$. Such
operations are classified by
$[K(\bZ,0),K(\bZ,3)]=[\bZ,K(\bZ,3)]=H^3(\bZ)=0$. So there is no
non-vanishing stable homology operations acting on these degrees, and
the spectral sequence stabilizes. There are no extension ambiguities
either in going from the filtration to the K-theory group, so we
conclude that $\tilde K_0(Y_5)=H_2(Y_5)=H^3(Y_5)$.

The relation between the reduced and non-reduced K-homology groups is
$K_0(Y_5)=\bZ\oplus \tilde K_0(Y_5)$ (see for instance eq.~(1.5) in
\cite{Ruffino:2009zp}), so $K_0(Y_5)=\bZ\oplus \tilde K_0(Y_5)$. Note
also that $Y_5$ admits a Spin structure (since its normal bundle in
the Calabi-Yau cone $X_6$ is trivial, and $X_6$ is Spin), and in
particular a Spin$^c$ structure, or in other words it is
K-orientable. So we can apply Poincaré duality, and
$K_0(Y_5)=K^1(Y_5)$. We conclude that
\begin{equation}
  K^1(Y_5) = H^3(Y_5)\oplus H^5(Y_5)
\end{equation}
and similarly
\begin{equation}
  K^0(Y_5) = H^0(Y_5)\oplus H^2(Y_5)\, .
\end{equation}

It is also clear that the K-theory groups of $\cM_4$ agree with the formal sums of cohomology
groups, since by the Chern isomorphism
$K^i(\cM_4)\otimes \bQ \cong \bigoplus_{n} H^{2n+i}(\cM_4;\bQ)$, and
we are assuming that $\cM_4$ has no torsion, so the relevant
Atiyah-Hirzebruch spectral sequence has no non-vanishing
differentials.

Finally, we can use the Künneth exact sequence in K-theory
\cite{ATIYAH1962245}
\begin{equation} \label{eqn:Kunnethformula}
  0 \to \bigoplus_{i+j=m} K^i(X)\otimes K^j(Y) \to K^m(X\times Y) \to \bigoplus_{i+j=m+1} \Tor_\bZ(K^i(X), K^j(Y)) \to 0
\end{equation}
to assemble these results together, and prove the statement in
footnote~\ref{fn:K-theory} that we can use ordinary cohomology for
classifying IIB flux in these backgrounds.

\bibliographystyle{JHEP}
\bibliography{refs}

\end{document}